\documentclass[prd,a4paper,preprint,preprintnumbers,nofootinbib]{revtex4-2}

\usepackage[a4paper, hdivide={1.91cm,,1.165cm}, vdivide={1.83cm,,3.0cm}]{geometry}

\usepackage{amsmath}
\usepackage{graphicx}
\usepackage{xspace}
\usepackage{color}
\usepackage{units}
\usepackage{slashed}
\usepackage[hyperfootnotes=false]{hyperref}
\hypersetup{linkcolor=red,
citecolor=green,
filecolor=cyan,
urlcolor=magenta}
\usepackage{gensymb}
\usepackage{booktabs}
\usepackage{multirow}
\usepackage{xcolor}
\usepackage{setspace}

\newcommand{\ra}{\rightarrow}

\newcommand{\beq}{\begin{eqnarray}}
\newcommand{\eeq}{\end{eqnarray}}
\newcommand{\Tcc}{T_{cc}^+}

\def\beq{\begin{equation}}
\def\eeq{\end{equation}}
\def\bea{\begin{eqnarray}}
\def\eea{\end{eqnarray}}

\def\nnb{\nonumber}

\def\nnb{\nonumber}
\def\la{\langle}
\def\ra{\rangle}

\def\qu{\la \bar u u \ra}
\def\qd{\la \bar d d \ra}

\def\gGgG{\la g^2 G^2 \ra}
\def\es{ &=& }
\def\ar{&+& }
\def\ek{&-& }
\def\cp{&\times&}

\begin{document}


\title{Determination of the spectroscopic parameters of beauty-partners of $T_{cc}$ from QCD}

\author{T.~M.~Aliev}
\email{taliev@metu.edu.tr}
\affiliation{Department of Physics, Middle East Technical University, Ankara, 06800, Turkey}

\author{S.~Bilmis}
\email{sbilmis@metu.edu.tr}
\affiliation{Department of Physics, Middle East Technical University, Ankara, 06800, Turkey}
\affiliation{TUBITAK ULAKBIM, Ankara, 06510, Turkey}

\author{M.~Savci}
\email{savci@metu.edu.tr}
\affiliation{Department of Physics, Middle East Technical University, Ankara, 06800, Turkey}

\date{\today}

\begin{abstract}
Motivated by the recent discovery of a new tetra-quark $\Tcc$ state with two charm quarks and two light quarks by LHCb collaboration, we calculate the spectroscopic parameters, namely, the mass and residues of beauty partners of $\Tcc$ within QCD sum rules. The obtained results are compared with the predictions of different approaches in the literature.

\end{abstract}

\maketitle

\newpage


\section{Introduction\label{intro}}
The study of hadron spectroscopy represents a promising way to understand the inner structure of the hadrons. The quantum chromodynamics (QCD) theory allows the existence of multiquark states in a color singlet state. The first observation of such states, $X(3872)$, was achieved by BELLE collaboration~\cite{Belle:2003nnu}. Many multi-quarks states have been discovered up to now. The properties of such states have been widely analyzed by theorists and experimentalists (see review~\cite{Chen:2016qju,Guo:2017jvc,Yamaguchi:2019vea}). 

All the discovered exotic hadrons have been achieved by analyzing their strong interactions. The measured decay widths lie in the domain of a few to a few hundred MeV. On the other hand, observing a long-lived exotic state that is stable with respect to the strong interaction would be exciting. The first candidates of such states are expected to be hadrons with two heavy quarks (Q) and two light antiquarks (q). Recently, LHCb collaboration has reported discovering such a state, namely tetraquark state $\Tcc$ with two charm quarks and two light antiquarks in the $D^0 D^0 \pi^+$ spectrum~\cite{LHCb:2021vvq,LHCb:2021auc} with quantum numbers $J^P = 1^+$. The observed decay width and the mass difference is~.\cite{LHCb:2021vvq} 
\begin{equation}
\begin{split}
\delta m &= m_{\Tcc} - (m_{D^{*+}} + m_{D^{0}}) = - 273 \pm 61 \pm 5^{+11}_{-14}~keV \\
\Gamma_{\Tcc} &= 410 \pm 165 \pm 43^{+18}_{-38}~keV~.
\end{split}
\end{equation}
This observation can stimulate studies of exotic hadrons in new ways. One of the possible interpretations of $\Tcc$ is that it is a compact $DD^*$ hadronic state. This discovery inspires us to use the QCD sum rules approach to investigate the mass and residues of all the exotic $T_{bbqq}$ states, potentially to be observed in the near future, with quantum numbers $J^{P} = 1^{+}$. In calculations, we also take into account the $SU(3)$ violation effects.

The paper is organized as follows. In Section~\ref{sec:2}, we derive sum rules for the mass and residue of the possible bottomonium states. Section~\ref{sec:3} is devoted to the numerical analysis of the obtained sum rules. This section also contains discussions and comparisons of our findings with the other approaches. And finally, we conclude our study.
\section{Mass sum rules for bottomonium states with $J^{P} = 1^+$}
\label{sec:2}
Before deriving the sum rules, we present all the possible bottomonium states with quantum numbers $J^{P} = 1^+$ (see Table ~\ref{tab:1}.)


\begin{table*}[t]
  \centering
  \renewcommand{\arraystretch}{1.4}
  \setlength{\tabcolsep}{7pt}
  \begin{tabular}{ccccc}
    \toprule
     $T_{bb}$ States             & Tetraquark Contents         \\
    \midrule
    $T_{bb}^{- -}$  & $B^- B^{*-}$ \\
    $T_{bb}^{-}$    & $B^0 B^{*-}$, $B^{-} B^{*0}$ \\
    $T_{bb}^{0}$    & $B^0 B^{*0}$ \\
    $T_{bb(s)}^{-}$ & $B_s^0 B^{*-}$, $B_s^{*0} B^{-}$ \\
    $T_{bb(s)}^{0}$ & $B_s^0 B^{*0}$, $B^{*0} B_s^{*0}$ \\
    $T_{bb(ss)}^{0}$ & $B_s^0 B_s^*$ \\
    \bottomrule
  \end{tabular}
  \caption{Possible states of $T_{bb}$ with $J^{P} = 1 ^{+}$.}
  \label{tab:1}
\end{table*}

We choose the interpolating current for $T_{bb}$ state with quantum numbers $J^P = 1^+$ as
\begin{equation}
    j_{\mu} = (\bar{q_1}^a i \gamma_5 Q_1^a) (\bar{q_2}^b \gamma_\mu Q_2^b)
\end{equation}
where $a$ and $b$ are the color indices, $q_1(q_2)$ is the light quark, and $Q_1 (Q_2)$ represents heavy quarks. The light and heavy quarks can either be the same or different. 

To obtain the mass sum rules we consider the following correlation function,
\begin{equation}
\begin{split}
    \Pi_{\mu \nu} &= i \int d^4 x \langle 0 | T \big( j_\mu(x) j_{\nu}^{\dagger}(0) \big) | 0 \rangle \\
                  &= (-g_{\mu \nu} + \frac{p_\mu p_\nu}{p^2}) \Pi_1 + \frac{p_\mu p_\nu}{p^2} \Pi_2    
\end{split}
\end{equation}
where $\Pi_1$ and $\Pi_2$ represent the invariant functions corresponding to transversal and longitudinal parts of the correlation function. 

According to the QCD sum rules method, the correlation function is calculated in two different kinematical regions; in terms of hadrons and in the deep Euclidean region $p^2 \rightarrow - \infty$ using the Operator product expansion(OPE). Matching these two representations and performing Borel transformation with respect to the variable $-p^2$ to suppress the contributions from the excited states and continuum as well as to enhance the ground state contribution, we get mass sum rules.

Now let us calculate the correlation function from the hadronic side first. The hadronic representation of the correlation function satisfies the dispersion relation,
\begin{equation}
    \Pi_i = \int \frac{\rho_i(s) ds }{s-p^2} + \text{subtractions} ,
\end{equation}
where $i = 1(2)$ corresponds to $\Pi_1$ and $\Pi_2$ and $\rho_i(s)$ is the corresponding spectral density.
%
Inserting a complete set of intermediate states with the same quantum numbers as the interpolating current $J^P = 1^+$ and isolating the ground state contribution for the spectral density we get 
\begin{equation}
    \rho_i = \lambda_i^2 \delta(s - m_T^2) + ...
\end{equation}
where dots denotes the contribution from higher states. To obtain this spectral density, we used the standard definition 
\begin{equation}
    \langle 0 | j_\mu | T_{QQ}(p) \rangle = \lambda \epsilon_\mu~,
\end{equation}
where $\lambda$, and $\epsilon_\mu$ are the decay constant, and vector polarization of $T_{QQ}$ bottomed state. Then we performed summation over spins of $T_{QQ}$ and separated the coefficient of the structure $g_{\mu \nu} - \frac{p_\mu p_\nu}{p^2}$.

As we already noted that the theoretical part of the correlation function is calculated in the deep Euclidean domain with the help of OPE. After simple calculations, we obtain the correlation function for $T_{bb}$  as follows:
for case $Q_1 = Q_2 = Q$ and $q_1 \neq q_2$:
\begin{equation}
  \label{eq:7}
\begin{split}
        \Pi_{\mu \nu} = i \int d^4 x e^{i p x} \Bigg\{ & 
        Tr \big[S_{Q}^{b a^\prime}(x) \gamma_5 S_{q_1}^{a^\prime a}(-x) \gamma_5 S_{Q}^{a b^\prime}(x) \gamma_\nu S_{q_2}^{b^\prime}(-x) \gamma_\mu \big] \\
        -& Tr\big[ \gamma_5 S_{Q_1}^{a a^\prime}(x) \gamma_5 S_{Q_1}^{a^\prime a}(-x) \big]  Tr\big[\gamma_\mu S_{Q}^{b b^\prime}(x) \gamma_\nu S_{q_2}^{b^\prime b}(-x) \big]
        \Bigg\}
\end{split}
\end{equation}
and for $Q_1 = Q_2 = Q$ and $q_1 = q_2 = q$ case:
\begin{equation}
  \label{eq:2}
  \begin{split}
\Pi_{\mu \nu}    = i \int d^4 x e^{i p x} \Bigg\{ & Tr \big[ \gamma_5 S_{Q}^{a b^\prime}(x) \gamma_\nu S_{q}^{b^\prime a}(-x) \big] Tr \big[ \gamma_\mu S_{Q}^{b a^\prime}(x) \gamma_5 S_{q}^{a b}(-x) \big] \\
   + & Tr[\gamma_5 S_{Q}^{a a^\prime}(x) \gamma_5 S_{q}^{a^\prime a}(-x)] Tr \big[\gamma_\mu S_{Q}^{b b^\prime}(x) \gamma_\nu S_{d}^{b b^\prime}(-x) \big] \\
   - & Tr \big[S_{Q}^{b a^\prime}(x) \gamma_5 S_{d}^{a^\prime a }(-x) \gamma_5 S_{Q}^{a b^\prime}(x) \gamma_\nu S_{q}^{b^\prime b}(-x) \gamma_\mu \big] \\
   - & Tr\big[ S_{Q}^{b b^\prime}(x) \gamma_\nu S_{d}^{b^\prime a}(-x) \gamma_5 S_{Q}^{a a^\prime}(x) \gamma_5 S_{q}^{a^\prime b}(-x) \gamma_\mu \big] \Bigg\}
  \end{split}
\end{equation}
The QCD sum rules are obtained by choosing the same Lorentz structures from both representations of the correlation function and then matching the coefficients of these structures. In our calculations, we choose the structure $g_{\mu \nu}$ since it only contains the contributions of spin-1 particles.
From Eq.~\eqref{eq:7}, it follows that  light and heavy quark propagators are needed to calculate the correlation function. Light quark propagator in $x$-representation up to the first order of light quark mass is given as~.\cite{Ioffe:1983ju,Chiu:1986cf}
\begin{equation}
  \label{eq:1}
  \begin{split}
    S_q^{ab}(x) &= \frac{i \slashed{x} \delta^{ab}}{2 \pi^2 x^4} - \frac{m_q}{4 \pi^2 x^2} \delta^{ab} - \frac{1}{12} \langle \bar{q} q \rangle \delta^{ab} + \frac{g_s G^{n}_{\mu \nu} (\frac{\lambda^n}{2})^{ab}}{32 \pi^2 x^2} (i \sigma^{\mu \nu} \slashed{x} + i \slashed{x} \sigma^{\mu \nu}) + i m_q \frac{\bar{q} q}{48} \slashed{x} \delta^{ab} \\
    &+ \frac{1}{192} m_0^2 \langle \bar{q} q \rangle x^2 \delta^{ab} 
     - i \frac{m_q}{1152} m_0^2 \langle \bar{q} q \rangle x^2 \slashed{x} \delta^{ab} - \frac{4 \pi}{3^9 2^{10}} \langle \bar{q} q \rangle \langle \alpha_s G^2\rangle x^2 \delta^{ab} + ... 
  \end{split}
\end{equation}
Besides, the heavy-Quark propagator in x-representation is (see for example~\cite{Huang:2012ti}).
\begin{equation}
  \label{eq:8}
  \begin{split}
    S_Q^{ab}(x) &= \frac{m_Q^2 \delta^{ab}}{(2 \pi^2)} \bigg\{ i \slashed{x} \frac{K_2(m_Q\sqrt{-x^2})}{(\sqrt{-x^2})^2} + \frac{K_1(m_Q\sqrt{-x^2})}{\sqrt{-x^2}} \bigg\} \\
    &- \frac{m_Q g_s G^{n}_{\mu \nu} (\frac{\lambda^n}{2})^{ab}}{8 (2 \pi)^2} \bigg\{ i[ \sigma^{\mu \nu} \slashed{x} + \slashed{x} \sigma^{\mu \nu}] \frac{K_1(m_Q\sqrt{-x^2})}{\sqrt{-x^2}} + 2 \sigma^{\mu \nu} K_0 \bigg \}\\
    &- \frac{\delta^{ab} \langle g_s^2 G^2 \rangle}{576 (2 \pi)^2} \bigg\{ ( i \slashed{x}m_Q - 6) \frac{K_1 (m_Q \sqrt{-x^2})}{\sqrt{-x^2}} (-x^2) + m_Q(x^2)^2 \frac{K_2(m_Q\sqrt{-x^2})}{(\sqrt{-x^2})^2} \bigg\}
  \end{split}
\end{equation}
where $K_i$ are the modified Bessel functions of the second kind. 

Using the Eqs. \eqref{eq:1} and \eqref{eq:8} for the light and heavy quark propagators, after lengthy calculation we can find the spectral density $\rho_1$. The expressions of the spectral density are presented in Appendix.

Matching the expressions of spectral density $\rho_1(s)$ from hadronic and OPE parts and performing Borel transformation with respect to variable $-p^2$ to suppress the continuum and higher states as well as enhancing the contribution of the ground state by using the quark-hadron duality, we get the sum rules for the spectroscopic parameters (mass and residue) of the considered $T_{bb}$ meson:
\begin{equation}
\label{eq:9}
    \lambda^2  e^{-m_T^2/M^2} = \int^{s_0}_{s_{min}} ds \rho_1(s) e^{-s/M^2}
\end{equation}
where $s_{min} = (m_{Q_1}+m_{Q_2}+m_{q_1}+m_{q_2})^2 $,  $M^2$ and $s_0$ are the Borel mass parameter square and continuum threshold respectively. Getting derivative with respect to $(-1/M^2)$ and dividing the obtained result to Eq.~$\eqref{eq:9}$ we get the desired sum rule for the mass of $T_{bb}$

\begin{equation}
\label{eq:10}
    m_T^{2} = \frac{\int^{s_0}_{s_{min}} s \rho_1(s)  e^{-s/M^2} ds }{\int_{s_{min}}^{s_0} \rho_1(s) e^{-s/M^2} ds}
\end{equation}
Moreover, once the $m_{T}^2$ is obtained from Eq.~\eqref{eq:10}, we can obtain the residue by putting it into the Eq.~\eqref{eq:9}.

At the end of this section, we would like to note that next to leading order (NLO), perturbative corrections to the correlation function for tetraquark systems with two heavy quarks have been calculated in~,\cite{Tang:2019nwv} and it is obtained that the radiative corrections are quite small. Similarly, NLO corrections to the correlation function for the considered states are also expected to be small, and hence we neglected them in our calculations. 

\section{Numerical Analysis}
\label{sec:3}
Having derived the expressions for the mass and residue for $T_{bb}$, we present the numerical results of these quantities in this section.  The values of the input parameters used in our calculations are depicted in Table ~\ref{tab:2}. For the heavy quark masses, we used their $\overline{MS}$ values. 

\begin{table*}[t]
  \centering
  \renewcommand{\arraystretch}{1.4}
  \setlength{\tabcolsep}{7pt}
  \begin{tabular}{ccccc}
    \toprule
     Parameters             & Value         \\
    \midrule
    $\overline{m}_s(1~GeV)$ & $0.114 \pm 0.021~GeV$~\cite{PhysRevD.98.030001}\\ 
    $\overline{m}_b(\overline{m}_b)$ & $4.18 \pm 0.03~GeV$~\cite{PhysRevD.98.030001}  \\ 
    $m_K$    & $0.497~GeV$~\cite{PhysRevD.98.030001}  \\ 
    $\langle \bar{q} q \rangle$    & $-(0.245)^3~GeV^3$~\cite{PhysRevD.98.030001}  \\
    $\langle \bar{s} s \rangle$    & $(0.8 \pm 0.2)~\langle \bar{q} q \rangle ~GeV^3$ ~\cite{Gelhausen:2014jea}\\
    $m_0^2$    & $(0.8 \pm 0.2)~GeV^2$~\cite{Ioffe:1981kw} \\
    $ \langle \frac{\alpha_s G^2}{\pi} \rangle $    & $(0.012^{+0.006}_{-0.0012})~GeV^4$ ~\cite{Ioffe:2002ee}, \\
    \bottomrule
  \end{tabular}
  \caption{The values of the input parameters used in our calculations.}
  \label{tab:2}
\end{table*}
The Borel mass parameters, $M^2$, and continuum threshold, $s_0$, are the auxiliary parameters of the sum rules. Hence, physical quantities are expected to be independent of these parameters. For this purpose, we need to find the ``working regions" of these parameters where measurable quantities are insensitive to the variation of them.
The upper bound of $M^2$ is determined from the condition that the higher states and continuum contributions should be suppressed compared to the total result, i.e., the dominance of pole contribution. The pole contribution (PC) is defined as
\begin{equation}
  \label{eq:pole}
  PC(s_0,M^2) = \frac{\int_{s_{min}}^{s_{max}} \rho_1(s) e^{-s/M^2} ds}{\int_{s_{min}}^{\infty} \rho_1(s) e^{-s/M^2} ds}
\end{equation}
We require that the pole contribution constitute more than $50\%$ of the total result. The minimum value of the $M^2$ is obtained from the requirement that the operator product expansion  is convergent. For this goal, we consider the following quantity

\begin{equation}
  \label{eq:cond}
  \mathcal{D}(n) = \frac{\int_{s_{min}}^{s_0} \rho_1^{(n)}(s) e^{-s/M^2} ds}{\int_{s_{min}}^{s_0} \rho_1(s) e^{-s/M^2} ds}
\end{equation}
where $\rho_1^{(n)}$ corresponds to the spectral density describing the contribution of the condensate of dimension $n$. We demand that the contribution of condensate with maximum dimension in sum rules should be less than, say,  $5\%$ of the total result. 
The minimum value of $M^2$ is determined from the condition that the continuum and higher state contributions should be suppressed compared to the total result. Besides, the upper bound of $M^2$ is obtained from the requirement the operator product expansion should be convergent, i.e., the power corrections should be suppressed with respect to the total result. 
\begin{table*}[t]
  \small
  \centering
  \renewcommand{\arraystretch}{1.4}
  \setlength{\tabcolsep}{7pt}
  \begin{tabular}{lccccc}
    \toprule
    States       & $M^2(GeV^2)$ & $s_0 (GeV^2)$ & mass(GeV) & $\lambda^2 (GeV^{10})$ & Pole contribution $(\%)$\\ 
    \midrule
    $ BB^*$      & $10 - 20 $   & $115 - 120$  & $10.3 \pm 0.1$  & $0.030 \pm 0.008$ & $64 - 38$    \\     
    $ B_s B^*$   & $10 - 20 $   & $120 -125$   & $10.4 \pm 0.1$  & $0.05 \pm 0.01$   & $68 - 38$   \\    
    $ B B_s^*$   & $10 - 20 $   & $120-125 $   & $10.4 \pm 0.1$  & $0.05 \pm 0.01$   & $68 - 38$    \\   
    $ B_s B_s^*$ & $10 - 20 $   & $125-130 $   &$10.7 \pm 0.12$  &$0.07 \pm 0.02$   & $72 - 42$    \\  
    \bottomrule
  \end{tabular}
  \caption{The working regions of $M^2$ and $s_0$ as well as the value of residues for the considered states.}
  \label{tab:son}
\end{table*}

The threshold parameter $s_0$ is determined via the minimized variation of the $T_{bb}$ mass with respect to the variation of $M^2$. Considering these restrictions, one can determine the $s_0$ and $M^2$ for $T_{bb}$ states. Our numerical analysis shows that the conditions mentioned above are satisfied in the regions of $M^2$ and $s_0$ presented in Table ~\ref{tab:son}. In this Table, we also present the pole contribution. 

\begin{table*}[t]
  \small
  \centering
  \renewcommand{\arraystretch}{1.4}
  \setlength{\tabcolsep}{3.2pt}
  \begin{tabular}{lcccc}
    \toprule    
    States                   & $ BB^* (10604)$ & $ B_s B^*(10692)$ & $ B B_s^* (10695)$ & $ B_s B_s^*(10782)$ \\
    \midrule
    This Study               & -304            & -291.5            & -295.1             & -82                 \\
    \cite{Karliner:2017qjm}  & -215            & ---               & ---                & ---                 \\  
    \cite{Eichten:2017ffp}   & -121            & -48               & -48                & ---                 \\  
    \cite{Braaten:2020nwp}   & -133            & -47               & NB                 & NB                  \\  
    \cite{Cheng:2020wxa}     & -116            & -21               & -21                & NB                  \\  
    \cite{Dias:2011mi}       & -464            & ---               & ---                & ---                 \\  
    \cite{Navarra:2007yw}    & -404            & ---               & ---                & ---                 \\  
    \cite{Gao:2020ogo}       & -224            & -120              & -123               & ---                 \\  
    \cite{Ren:2021dsi}       & -6              & $\sim 0$          & -2.1               & $\sim 0$                 \\  
    \cite{Hernandez:2019eox} & -150            & ---               & ---                & ---                 \\  
    \cite{Yang:2019itm}      & -35             & ---               & ---                & ---                 \\
    \cite{Yang:2009zzp}      & -120.9          & ---               & ---                & ---                 \\
    \cite{Tan:2020ldi}       & -317.6          & ---               & ---                & ---                 \\  
    \cite{Lu:2020rog}        & -54             & ---               & ---                & ---                 \\
    \cite{Ebert:2007rn}      & -102            & NB                & NB                 & NB                  \\
    \cite{Meng:2020knc}      & -173            & -59               & -59                & ---                 \\
    \cite{Zhang:2021yul}     & NB              & NB                & NB                 & ---                 \\  
    \cite{Noh:2021lqs}       & -145            & -42               & ---                & ---                 \\  
    \cite{Deng:2018kly}      & -278            & -49               & -49                & NB                  \\  
    \cite{Francis:2016hui}   & -189            & -98               & -98                & ---                 \\  
    \cite{Junnarkar:2018twb} & -143            & -87               & -87                & ---                 \\  
    \cite{Leskovec:2019ioa}  & -128            & ---               & ---                & ---                 \\  
    \cite{Mohanta:2020eed}   & -167            & ---               & ---                & ---                 \\  
    \midrule
    \bottomrule
  \end{tabular}
  \caption{Theoretical results for the binding energy (in $\rm{MeV}$ units) of $T_{bb}$ states predicted by different approaches are presented. Here, NB stands for the no-bound state. The values inside the parenthesis correspond to the threshold values of the corresponding meson-meson states. }
  \label{tab:44}
\end{table*}

\begin{figure}[t]
\includegraphics[width=0.49\textwidth]{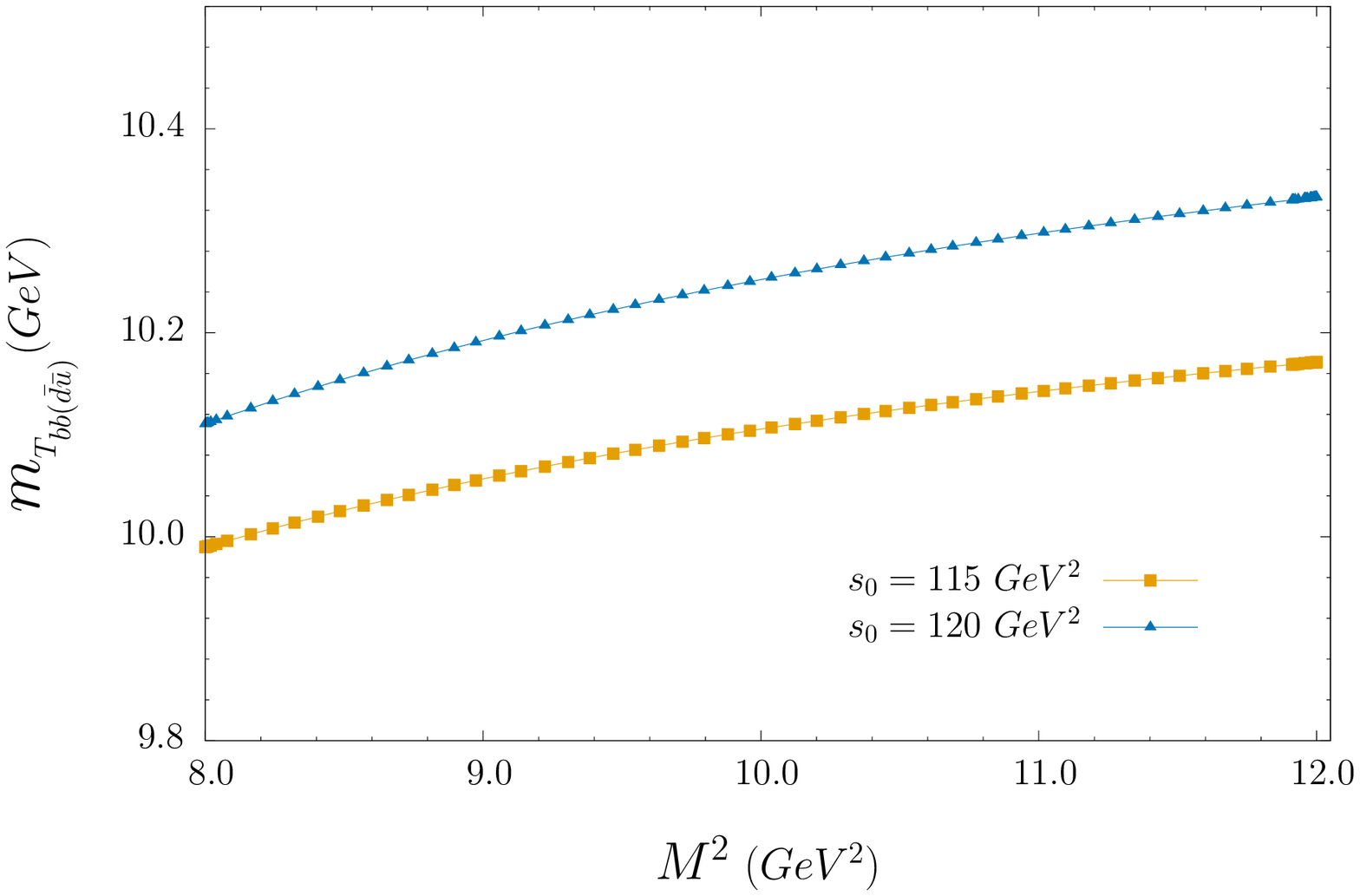}
\includegraphics[width=0.49\textwidth]{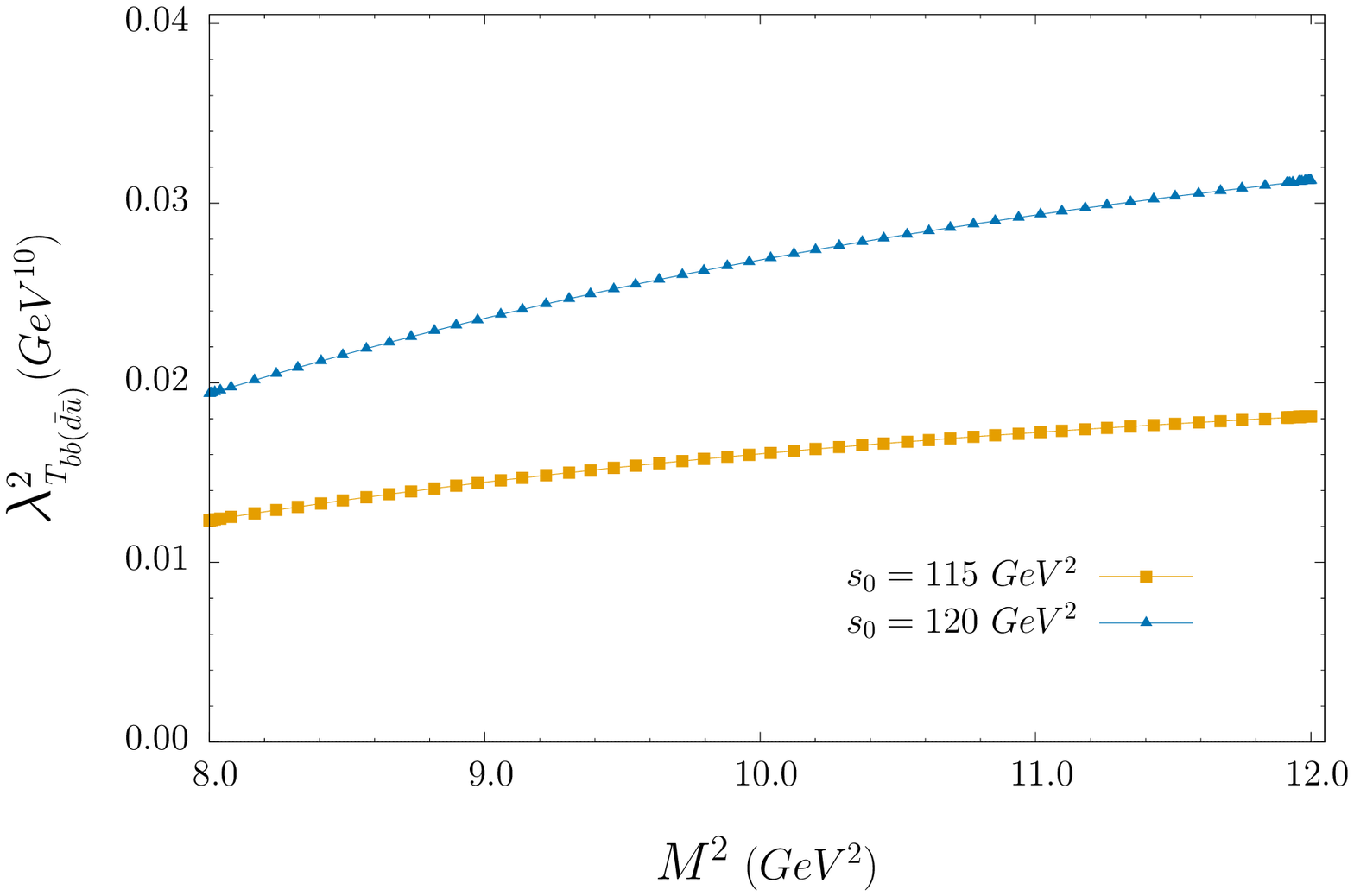}
\caption{
The dependencies of mass and residue of the $T_{bb}$ state on Borel mass square at the fixed values of $s_0$.
}
\label{fig:fig1}
\end{figure}

In Fig.~\ref{fig:fig1}, we presented the dependency of the mass of $T_{bb}$ state on Borel mass parameters $M^2$ at the fixed values of $s_0$. From this figure, we get $m = 10300 \pm 100$ MeV for the $T_{bb}$ state.  Performing similar numerical analysis for the other $T_{bb}$ states such as $B_s B^*(B B_s^*)$ and $B_s B_s^*$, we obtained the residues and mass values of the corresponding states that are presented in Table ~\ref{tab:son}. In the determination of the mass, we take into account the uncertainties coming from input parameters as well as from the variation of $M^2$ and $s_0$.
%
%
%
In Table~\ref{tab:44}, we present our theoretical results for the binding energy (in MeV) of $T_{bb}$ states. Moreover, for completeness in this Table, we also present the binding energy predicted by the other approaches. From the comparison, we observe that our predictions on the binding energy of $T_{bb}$ are larger than the results of other approaches, especially for $B_sB^*$, $BB_s^*$, and $B_sB_s^*$ states. On the other hand, for the $BB^*$ state, our finding on the mass difference is less than the predictions obtained in~\cite{Dias:2011mi,Navarra:2007yw,Tan:2020ldi} and larger than the other results. This circumstance hopefully can be clarified in future experiments. 

Finally, we would like to make the following remark. To derive the sum rules, we used a simple one-pole approximation. However, for the case of a multiquark state, the physical side of the sum rules in principle can receive contributions from two-particle hadronic states at least. This point was first raised in~\cite{Kondo:2004cr,Lee:2004xk}. The multiquark state can decay into kinematically allowed states; hence, we should take into account the decay width of multiquark states.
Once we do that, the quark propagator should be replaced in the following way.
\begin{equation}
    \frac{1}{m^2 - p^2} \rightarrow \frac{1}{m^2 - p^2 - i \Gamma \sqrt{p^2}  }~,
\end{equation}
where $\Gamma$ is the decay width of $T_{bb}$ state.
For the observed $\Tcc$ state, the width was measured as $\Gamma_{T_{cc}} \sim 400~keV$, which was smaller than the mass of $\Tcc$. We expect a similar case for the $T_{bb}$ state. In this situation, one can safely neglect the contributions of two hadronic states in the phenomenological part of the correlation function.

\section{Conclusion\label{conclusion}}
Motivated by the recent observation of double charmed tetraquark  $\Tcc$ by LHCb collaboration, we investigated the mass spectrum and residues of the possible doubly bottomed tetraquark states with quantum numbers $J^P = 1^+$ within the QCD sum rules method. We compared our results with the findings of other theoretical approaches. Future LHCb experiments will hopefully hunt the states considered in this study. 
%
\bibliographystyle{utcaps_mod}
\bibliography{mybib.bib}
%
%
%
%
%
\appendix*
\section{The expression of the spectral densities}
\label{app:formulas}
In this Appendix, we present the expressions of the spectral density for the $BB^*$ system with the quantum number $J^{P} = 1^+$.
The spectral density contains the perturbative contribution, the quark, and gluon condensate terms, quark-gluon mixed terms, four quark condensates, and dimension eight, nine terms. Their expressions are;
\bea
\rho^{(pert)} \es
   {1\over (3 \times 2^{14}) \pi^6}
\int_{\alpha_{min}}^{\alpha_{max}} {d\alpha \over \alpha^3}
 \int_{\beta_{min}}^{\beta_{max}} {d\beta \over \beta^3}
 \Big[ (\alpha + \beta) m_b^2 - \alpha \beta s \Big]^2 \nnb \\
\cp   \Big\{ (1 - \beta) \beta m_b^3 \big[ (8 + 29 \beta (1 + \beta)) m_b -
      120 \beta (m_d + \beta m_d + 2 m_u) \big] \nnb \\
\ek \alpha^4 (29 m_b^2 - 33 \beta s)
     (m_b^2 - \beta s) + 2 \alpha m_b \big\{  m_b \big[ (4 + 21 \beta - 58 \beta^3) m_b^2 \nnb \\
\ar 12 \beta (33 - 32 \beta) m_d m_u - 60 \beta m_b (m_d - 3 \beta^2 m_d + 2 m_u -
          4 \beta m_u) \big] \nnb \\
\ek (1 - \beta) \beta \big[ (4 + 31 \beta (1 + \beta)) m_b -
        60 \beta (m_d + \beta m_d + 2 m_u) \big]  s \big\} \nnb \\
\ek    2 \alpha^3 \beta (58 m_b^4 - 60 m_b^3 m_d - 93 \beta m_b^2 s + 60 \beta m_b m_d s +
      33 \beta^2 s^2) \nnb \\
\ar 3 \alpha^2 \big\{ (7 - 58 \beta^2) m_b^4 +
      40 \beta m_b^3 (3 \beta m_d + 2 m_u) - 80 \beta^2 m_b (\beta m_d + m_u) s \nnb \\
\ek 2 \beta m_b^2 (128 m_d m_u + 9 s - 31 \beta^2 s) -
      \beta^2 s \big[ 8 m_d m_u - 11 (1 - \beta^2) s \big] \big\} \Big\} 
\eea
\bea
\rho^{(\qu)} \es
 - {\qu \over 2^{10} \pi^4}
\int_{\alpha_{min}}^{\alpha_{max}} {d\alpha \over \alpha (1-\alpha)}
\big[ m_b^2 - (1 - \alpha) \alpha s \big] \nnb \\
\cp \big[  20 (1 - \alpha) m_b m_d m_u -
    m_b^2 (2 m_d + 11 m_u) + (1 - \alpha) \alpha (2 m_d + 11 m_u) s \big] \nnb \\
\ek {\qu \over 2^{10} \pi^4}
\int_{\alpha_{min}}^{\alpha_{max}} {d\alpha \over \alpha^2}
 \int_{\beta_{min}}^{\beta_{max}} {d\beta \over \beta}
 \big[ (\alpha + \beta) m_b^2 - \alpha \beta s \big] \nnb \\
\cp  \Big\{m_b \big\{ 4 m_b \big[ 5 (\alpha + \beta) m_b - 11 \alpha m_d \big] +
      \alpha \big[ 9 (\alpha + \beta) m_b - 20 \beta m_d \big] m_u \big\} \nnb \\
\ek \alpha \beta (20 m_b + 11 \alpha m_u) s \Big\}
\eea
%
\bea
\rho^{(\qd)} \es
  {\qd \over 2^{10} \pi^4}
\int_{\alpha_{min}}^{\alpha_{max}} {d\alpha \over \alpha (1-\alpha)}
 \big[m_b^2 - (1 - \alpha) \alpha s \big]
   \Big\{ m_b \big\{ 11 m_b m_d \nnb \\
\ar 2 \big[ m_b - 10 (1 - \alpha) m_d \big] m_u \big\} -
    (1 - \alpha) \alpha (11 m_d + 2 m_u) s) \nnb \\
\ek {\qd \over 2^{10} \pi^4}
\int_{\alpha_{min}}^{\alpha_{max}} {d\alpha \over \alpha^2}
 \int_{\beta_{min}}^{\beta_{max}} {d\beta \over \beta}
   \big[ (\alpha + \beta) m_b^2 - \alpha \beta s \big]
   \Big\{ m_b^2 \big\{ (\alpha + \beta) \nnb \\
\cp \big[ 20 (\alpha + \beta) m_b + 9 \alpha m_d \big] - 44 \alpha m_u \big\} -
    \alpha \beta \big[ 20 (\alpha + \beta) m_b + 11 \alpha m_d \big] s \Big\}
\eea
%
\bea
\rho^{(\gGgG)} \es
 {\gGgG \over (3 \times 2^{15}) m_b \pi^6}
\int_{\alpha_{min}}^{\alpha_{max}} {d\alpha \over \alpha (1-\alpha)}
      \Big\{ \big[m_b^2 - (1 - \alpha) \alpha s \big] \nnb \\
\cp \Big( m_b \big[ 33 m_b^2 + 80 (1 - \alpha) m_d m_u - 
20 m_b (m_d + m_u) \big] \nnb \\
\ek (1 - \alpha) \alpha \big[ 33 m_b - 20 (m_d + m_u) \big] s \Big) \Big\} \nnb \\
\ek {\gGgG \over (3 \times 2^{16}) m_b \pi^6}
\int_{\alpha_{min}}^{\alpha_{max}} {d\alpha \over \alpha^2}
 \int_{\beta_{min}}^{\beta_{max}} {d\beta \over \beta^2}
\Big\{
     \beta m_b^4 \Big( \big\{108 + \beta \big[ 219 + \beta (360 + 233 \beta) \big] \big\} m_b \nnb \\
\ek      24 \beta \big\{ \big[-6 + 4 (-1 + \beta) \beta \big] m_d + (-3 + 5 \beta) m_u \big\} \Big) \nnb \\
\ek 3 \alpha^4 m_b (m_b^2 - \beta s)^2 + 2 \alpha^3 (m_b^2 - \beta s)
     \big\{2 m_b^2 \big[ (27 + 47 \beta) m_b + 18 \beta m_d \big] \nnb \\
\ar \beta^2 (-61 m_b + 80 m_d) s \big\}  \nnb \\
\ar 2 m_b^2 \alpha \Big( m_b \Big[ 3 \big\{18 + \beta \big[1 + 2 \beta (82 + 55 \beta) \big] \big\} m_b^2 +
        6 \beta (12 + 35 \beta) m_d m_u \nnb \\
\ek  4 \beta m_b \big\{ 3 \big[-6 + 5 \beta (14 + \beta) \big] m_d +
          (-9 + 94 \beta) m_u \big\} \Big] \nnb \\
\ek \beta \Big[ \big\{ 54 + \beta \big[111 + 5 \beta (114 + 37 \beta) \big] \big\} m_b \nnb \\
\ek        12 \beta \big\{ \big[-6 + \beta (-4 + 9 \beta) \big] m_d + (-3 + 10 \beta) m_u \big\} \Big] s \Big)\nnb \\
\ar \alpha^2 \Big[ m_b^3 \Big( 3 (-71 + 244 \beta + 206 \beta^2) m_b^2 +
        8 \beta m_b \big[ 6 (-37 + \beta) m_d - 79 m_u \big] + 420 \beta m_d m_u \Big) \nnb \\
\ar      2 \beta m_b \Big( m_b \big\{ \big[105 - \beta (702 + 343 \beta) \big] m_b + 8 \beta (111 + 19 \beta)
           m_d \big\} \nnb \\
\ar 94 \beta (4 m_b - 3 m_d) m_u \Big) s +
      \beta^2 \big\{ \big[ 3 + \beta (780 + 137 \beta) \big] m_b - 120 \beta (\beta m_d + m_u) \big\} s^2 \Big]
      \Big\} 
\eea
%
\bea
\rho^{(m_0^2 \qu)} \es
-  { m_0^2 \qu \over (3 \times 2^{10}) \pi^4}
\int_{\alpha_{min}}^{\alpha_{max}} {d\alpha \over \alpha}
\Big\{ m_b \big[ 30 m_b^2 + 10 m_d m_u (1 - \alpha) \alpha \nnb \\
\ek  \alpha m_b (39 m_d  + 6 \alpha m_d + 12 m_u + 22 \alpha m_u) \big] \nnb \\
\ek  (1 - \alpha) \alpha (30 m_b - 9 \alpha m_d - 22 \alpha m_u) s \Big\} \nnb \\
\ek { m_b^2 m_u m_0^2 \qu \over (3 \times 2^{10}) \pi^4}
\int_{\alpha_{min}}^{\alpha_{max}} 
d\alpha \int_{\beta_{min}}^{\beta_{max}} d\beta 
\eea
\bea
\rho^{(m_0^2 \qd)} \es 
 - {m_0^2 \qd  \over (3 \times 2^{10}) \pi^4}
\int_{\alpha_{min}}^{\alpha_{max}} {d\alpha \over \alpha}
\Big\{ m_b \big[ 30 m_b^2 + 20 m_d m_u (1 - \alpha) \alpha \nnb \\
\ek  \alpha m_b (12 m_d + 22 \alpha m_d + 39 m_u + 6 \alpha m_u) \big] \nnb \\
\ek  (1 - \alpha) \alpha (30 m_b - 22 \alpha m_d - 9 \alpha m_u) s \Big\} \nnb \\
\ar {m_0^2 \qd \over (3 \times 2^{10}) \pi^4}
\int_{\alpha_{min}}^{\alpha_{max}} {d\alpha \over \alpha} \int_{\beta_{min}}^{\beta_{max}} d\beta
\Big\{ m_b^2 \big[30 (\alpha + \beta) m_b - \alpha m_d
\big] - 30 m_b \alpha \beta s \Big\} 
\eea
\bea
\rho^{(\qd \qu)} \es {\qd \qu \over (3 \times 2^8) \pi^2}
\int_{\alpha_{min}}^{\alpha_{max}} d\alpha
\Big\{4 m_b \big[ 13 m_b - 5 (2 m_d + m_u) \big] \nnb \\
\ar \alpha \big[ 12 m_0^2 + 8 m_b (m_b + 5 m_d) + 
m_u (20 m_b + 33 m_d) - 12 s \big] \nnb \\
\ek   3 \alpha^2 (4 m_0^2 + 11 m_d m_u - 4 s) \Big\} 
\eea
\bea
\rho^{(\qu \gGgG)} \es
 {\qu \gGgG \over (3^2 \times 2^{13})  m_b \pi^4}
\int_{\alpha_{min}}^{\alpha_{max}} {d\alpha \over \alpha}
 \Big\{  4 \big[ 21 - 2 \alpha (17 - 5 \alpha) \big]  m_b^2 \nnb \\
\ek  \alpha m_b \big[ 2 (61 + 74 \alpha + 6 \alpha^2) m_d +
(133 - 130 \alpha) m_u \big] \nnb \\
\ar  60 (1 - \alpha)^2 \alpha (m_d m_u - 2 s) \Big\} \nnb \\
\ek {\qu \gGgG \over (3 \times 2^{13}) \pi^4}
\int_{\alpha_{min}}^{\alpha_{max}} {d\alpha \over \alpha}
\int_{\beta_{min}}^{\beta_{max}} d\beta
(12 m_b - \alpha m_u) 
\eea
\bea
\rho^{(\qd \gGgG)} \es
 - {\qd \gGgG \over (3^2\times 2^{13}) m_b \pi^4}
\int_{\alpha_{min}}^{\alpha_{max}} {d\alpha \over \alpha}
 \Big\{ 4 \big[ 3 - \alpha (193 - 15 \alpha) \big] m_b^2 \nnb \\
\ar    \alpha m_b \big[ (763 - 760 \alpha) m_d + 
2 (61 + 74 \alpha + 6 \alpha^2) m_u \big] \nnb \\
\ek    40 (1 - \alpha) \alpha \big[ 3 (1 - \alpha) 
m_d m_u - (3 - 2 \alpha) s \big] \Big\} \nnb \\
\ar {\qd \gGgG \over (3^2 \times2^{13}) m_b \pi^4}
\int_{\alpha_{min}}^{\alpha_{max}} {d\alpha \over \alpha}
\int_{\beta_{min}}^{\beta_{max}} d\beta
\Big\{ 4 \big[ 8 \alpha - 9 (3 + 2 \beta) \big] m_b^2 \nnb \\
\ar 3 \alpha m_b m_d + 120 \alpha \beta s  \Big\} 
\eea
\bea
\rho^{(\gGgG^2)} \es
 - {\gGgG^2 \over (3^3 \times 2^{18}) m_b^2 \pi^6}
\int_{\alpha_{min}}^{\alpha_{max}} {d\alpha \over \alpha}
    \Big\{  \big[192 - \alpha ( 4143 + 584 \alpha) \big] m_b^2 \nnb \\
\ar     132 \alpha \big[ 1 - 2 (3 - \alpha) \alpha \big] m_d m_u +
    20 \alpha m_b \big[ (31 + 74 \alpha) m_d + 11 (m_u + 2 \alpha m_u) \big] \nnb \\
\ek    8 (3 - \alpha) (1 - \alpha) \alpha s \Big\} \nnb \\
\ar {\gGgG^2 \over (3^3 \times 2^{17}) m_b^2 \pi^6}
\int_{\alpha_{min}}^{\alpha_{max}} {d\alpha \over \alpha}
 \int_{\beta_{min}}^{\beta_{max}} {d\beta \over \beta}
\Big\{ m_b \big[ 18 \alpha^2 m_b + 18 (22 - 9 \beta) \beta m_b \nnb \\
\ar 5 \alpha \beta (37 m_b - 24 m_d) \big] -
6 \alpha \beta (3 \alpha + 5 \beta ) s \Big\}
\eea
\bea
\rho^{(m_0^2 \qd \gGgG)} \es
{ 5 (m_0^2 \qd \gGgG)  \over (3 \times 2^{12}) m_b \pi^4}
\int_{\alpha_{min}}^{\alpha_{max}} d\alpha
(1 - \alpha)^2 
\eea
\bea
\rho^{(m_0^2 \qu \gGgG)} \es
{ 5 (m_0^2 \qu \gGgG) \over  (3 \times 2^{11})  m_b \pi^4}
\int_{\alpha_{min}}^{\alpha_{max}} d\alpha
(1 - \alpha)^2  
\eea
where
\bea 
\alpha_{min} \es {s - \sqrt{ s (s - 4 m_b^2)} \over 2 s}~, \nnb \\
\alpha_{max} \es {s + \sqrt{ s (s - 4 m_b^2)} \over 2 s}~, \nnb \\
\beta_{min}  \es {m_b^2 \alpha \over s \alpha -m_b^2}~, \nnb \\
\beta_{max}  \es 1-\alpha~. 
\eea

In these expressions, we take into account the light quark masses. The term for the spectral density of $B_s B^* (BB_s^*)$ is obtained from these results by replacing $u (d)$ with strange quark.


\end{document}